\documentclass{ws-ijmpe}

\begin{document}

\markboth{Alexis Diaz-Torres, Leandro Romero Gasques and Michael Wiescher}{Astrophysical S-factor for $^{16}$O+$^{16}$O within the molecular picture}

\catchline{}{}{}{}{}

\title{ASTROPHYSICAL S-FACTOR FOR $^{16}$O+$^{16}$O WITHIN THE ADIABATIC 
MOLECULAR PICTURE}

\author{ALEXIS DIAZ-TORRES}

\address{Department of Nuclear Physics, Research School of Physical Sciences and Engineering, Australian National University, Canberra, ACT 0200, Australia \\
alexis.diaz-torres@anu.edu.au}
\author{LEANDRO ROMERO GASQUES}

\address{Centro de Fisica Nuclear, Universidade de Lisboa, Av. Prof. Gama Pinto 2 \\
1649-003 Lisboa, Portugal \\
lgasques@cii.fc.ul.pt}

\author{MICHAEL WIESCHER}

\address{Department of Physics and Joint Institute for Nuclear Astrophysics (JINA), \\ University of Notre Dame, Notre Dame, IN 46556, USA \\
mwiesche@nd.edu}

\maketitle

\begin{history}
\received{(received date)}
\revised{(revised date)}
\end{history}

\begin{abstract}
The astrophysical S-factor for $^{16}$O + $^{16}$O is investigated 
within the adiabatic molecular picture. It very well explains the available 
experimental data. The collective radial mass causes a pronounced resonant 
structure in the S-factor excitation function, providing a motivation for 
measuring the $^{16}$O + $^{16}$O fusion cross section at deep sub-barrier 
energies. 

\end{abstract}

\section{Introduction}

The fusion cross section $\sigma_{fus}$ at very low energies of reactions involving 
$^{12}$C and $^{16}$O is a crucial ingredient to calculate astrophysical reaction 
rates for different stellar burning scenarios in massive stars, in which 
$^{16}$O + $^{16}$O is the key reaction for the later oxygen burning phase. 
This cross section is usually represented by the 
S-factor ($S =\sigma_{fus} E e^{2\pi \eta}$, where $\eta$ is the Sommerfeld
parameter), as it facilitates the extrapolation of relatively high-energy fusion data because direct experiments at very low energies are very difficult to carry out. Unfortunately, there is a huge uncertainty in the S-factor resulted from the extrapolation 
of different phenomenological parametrizations that explain the high-energy data, as shown by Jiang et al.\cite{Jiang} for $^{16}$O + $^{16}$O. For $^{12}$C + $^{12}$C, the presence of pronounced molecular resonance structures makes it much more uncertain\cite{Aguilera}. 
These extrapolated values result in reactions rates that differ by many orders of magnitude\cite{Jiang}. Therefore, a direct calculation of the S-factor at energies of astrophysical interest ($< 3$ MeV) is essential. 

We report on an investigation\cite{Alexis1} of the fusion reaction $^{16}$O + $^{16}$O within the adiabatic molecular picture\cite{Greiner}, which is realistic at low incident energies. This is because the radial motion of the nuclei is expected to be adiabatically slow compared to the rearrangement of the two-center mean field of nucleons. In this reaction the nuclei are spherical and coupled channels effects are expected to be insignificant (the first collective excited state ($3^{-}$) of $^{16}$O is at 6.1 MeV), making its theoretical description relatively simple. Furthermore, abundant experimental data\cite{Exp} exist for comparison to the model calculations.

A basic microscopical model to describe the studied reaction is the two-center shell model (TCSM), a great concept introduced (in practice) in heavy-ion physics by the Frankfurt school\cite{TCSM}. We have used a new TCSM\cite{Alexis2} based on realistic Woods-Saxon (WS) potentials. The parameters of the asymptotic WS potentials including the
spin-orbit term reproduce the experimental single-particle energy
levels around the Fermi surface of $^{16}$O\cite{Alexis2},
whereas for $^{32}$S the parameters of the global WS potential by
Soloviev\cite{Soloviev} are used, its depth being adjusted to
reproduce the experimental single-particle separation energies\cite{Audi}. 
To describe fusion, the
potential parameters (including those of the Coulomb potential for protons) 
have to be interpolated between their values
for the separated nuclei and the compound nucleus. The parameters
can be correlated\cite{Alexis2} by conserving the volume enclosed by certain
equipotential surface of the two-center potential for all
separations $R$ between the nuclei.

\section{Calculations and discusion}

The adiabatic collective potential energy
surface $V(R)$ is obtained with Strutinsky's
macroscopic-microscopic method, whilst the radial dependent
collective mass parameter $M(R)$ is calculated with the cranking
mass formula\cite{Inglis}. For simplicity, the pairing contribution 
to the collective potential and radial mass is neglected. 
The rotational moment of inertia of
the dinuclear system is defined as the product of the cranking
mass and the square of the internuclear distance. The macroscopic
part of the potential results from the finite-range liquid drop
model\cite{Moeller1981} and the nuclear shapes of the TCSM\cite{Alexis2}. 
The microscopic shell corrections to the
potential are calculated with a novel method\cite{LanczosDiaz}. 
The TCSM is used to calculate the neutron and
proton energy levels\cite{Alexis1} $E_i$ as a function of the separation $R$
between the nuclei along with the radial coupling\cite{Alexis2}
between these levels that appears in the numerator of the cranking
mass expression,

\begin{equation}
M(R) = 2 \hbar^2 \sum_{i=1}^{A} \sum_{j > A}
\frac{|\langle j| \partial / \partial R |i \rangle|^2}
{E_j - E_i}.
\label{eq1}
\end{equation}

\begin{figure}
\begin{tabular}{cc}
\includegraphics[width=6.2cm,angle=0]{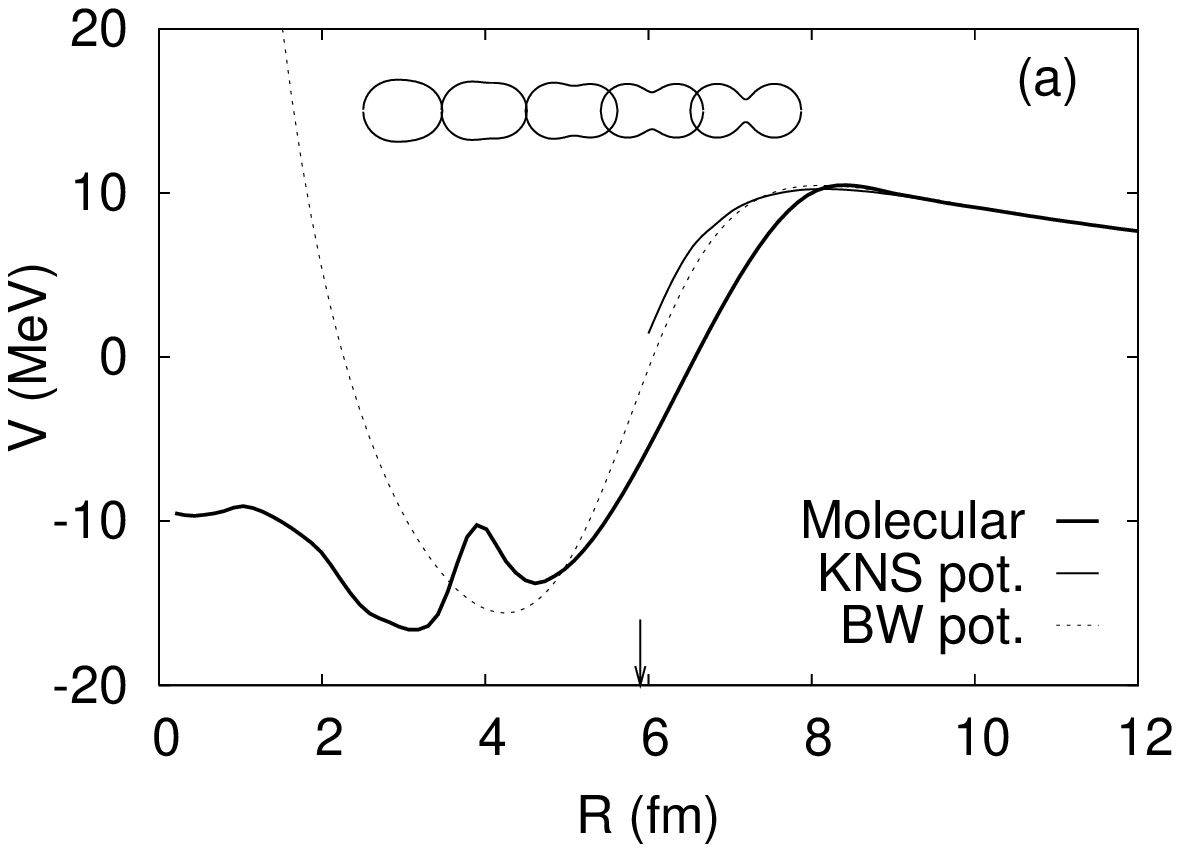} &
\includegraphics[width=6.2cm,angle=0]{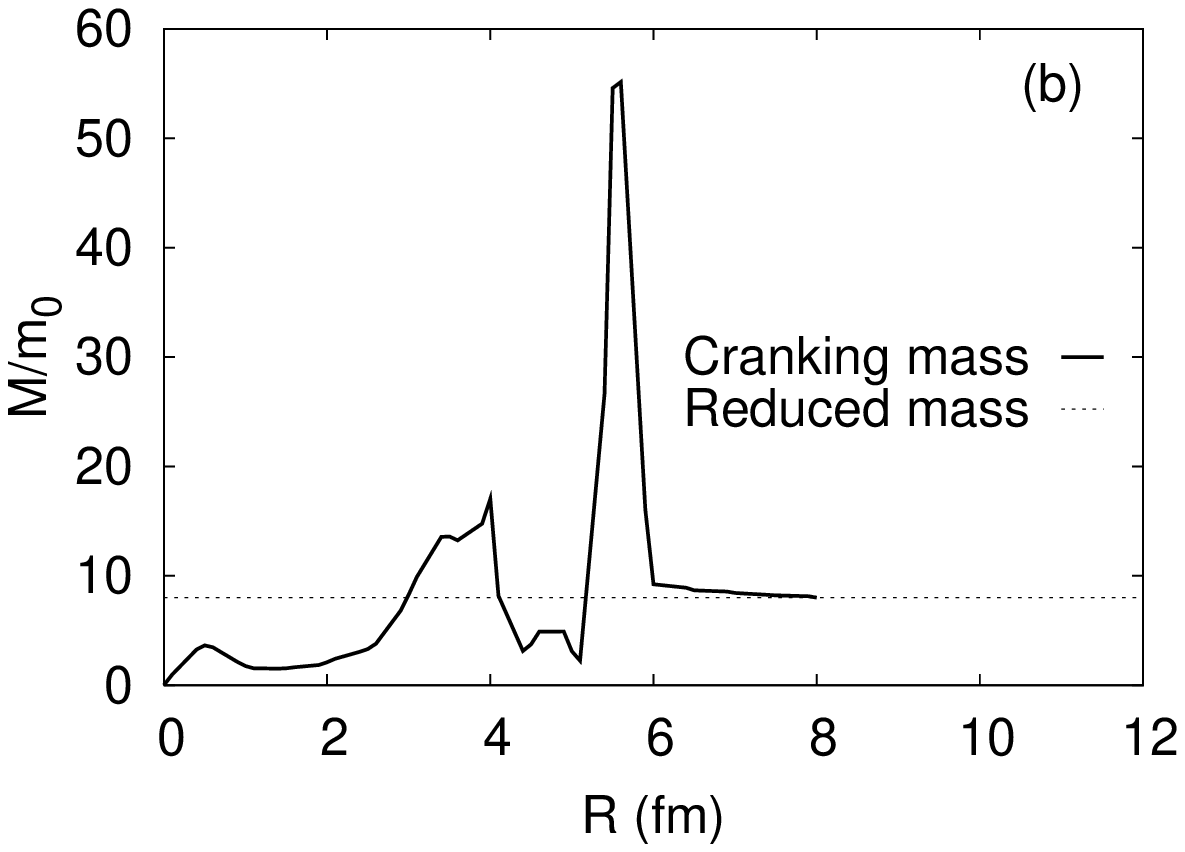}
\end{tabular}
\caption{(a) The s-wave collective potential energy as a function
of the separation between the nuclei for $^{16}$O + $^{16}$O. The
arrow indicates the geometrical contact separation. (b) The radial
dependent collective mass parameter (in units of nucleon mass
$m_0$). See text for further details.} \label{POTINERT}
\end{figure}

Figure \ref{POTINERT}a shows the s-wave molecular adiabatic
potential (thick solid curve) as a function of the internuclear
distance, which is normalized with the experimental $Q$-value of
the reaction ($Q=16.54$ MeV). The sequence of nuclear shapes
related to this potential\cite{Alexis2} is also presented. For
comparison we show the Krappe-Nix-Sierk (KNS) potential\cite{BW}
(thin solid curve) and the empirical Broglia-Winther (BW)
potential\cite{BW} (dotted curve). Effects of neck between the
interacting nuclei, before they reach the geometrical contact
separation (arrow), are not incorporated into the KNS potential. The
concept of nuclear shapes is not embedded in the BW potential
which tends to be similar to the KNS potential. Comparing the 
KNS potential to the molecular adiabatic potential we note that the neck
formation substantially decreases the potential energy after
passing the barrier radius ($R_b = 8.4$ fm). It will be shown that
the inclusion of neck effects is crucial to successfully explain the 
available S-factor data\cite{Exp} for the studied reaction.

Figure \ref{POTINERT}b shows the radial dependent cranking mass 
(thick solid curve), whilst the asymptotic reduced mass is
indicated by the dotted line. Just passing the barrier radius,
when the neck between the nuclei starts to develop, the cranking
mass slightly increases compared to the reduced mass and
pronounced peaks appear inside the geometrical contact separation.
For the studied reaction, these peaks are mainly caused by the
strong change of the single-particle wave functions during the
rearrangement of the shell structure of the asymptotic nuclei into
the shell structure of the compound system. 
In general, the peaks could also be due to avoided
crossings\cite{Alexis2} between the adiabatic molecular 
single-particle states\cite{Alexis1}, which can make 
the denominator of the cranking mass expression (\ref{eq1}) 
very small.
It is important to stress that the amplitude of these peaks may be 
reduced by (i) the pairing correlation that spreads out the single-particle 
occupation numbers around the Fermi surface, and (ii) the diabatic 
single-particle motion\cite{Alexis2} at avoided crossings, which can change those populations. 
For the strongest peak in Fig. \ref{POTINERT}b, which is located very close to 
the internal turning point for a wide range of sub-barrier energies, only aspect (i) 
may be relevant as the radial velocity of the nuclei is rather small there, suppressing 
the Landau-Zener transitions. For compact shapes, aspect (ii) may lead to intrinsic excitation of the composite system, but this is not important here for the calculation of the fusion cross section. Fusion is determined by the tunneling probability of the external Coulomb barrier, as explained below. 

\begin{figure}
\begin{center}
\includegraphics[width=8.0cm]{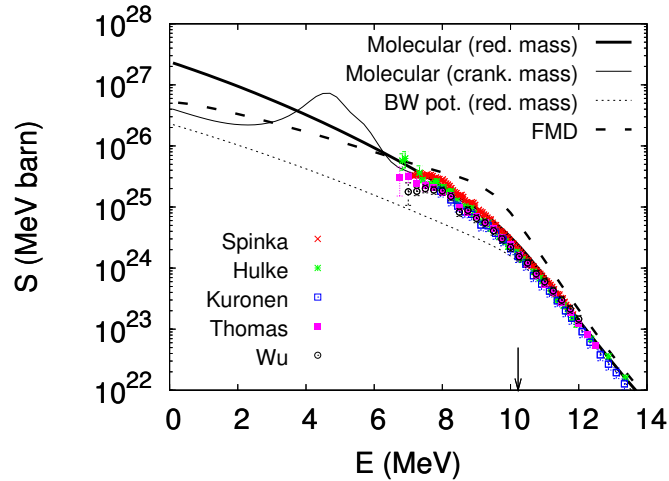}
\end{center}
\caption{(Color online) The S-factor as a function of the
center-of-mass energy for $^{16}$O + $^{16}$O. The curves are
theoretical calculations, whilst the symbols refer to experimental
data. The arrow indicates the Coulomb barrier of the molecular
potential of Fig \ref{POTINERT}a. See text for further details.}
\label{SFACT}
\end{figure}

Having the adiabatic potential and the adiabatic mass parameter,
the radial Schr\"odinger equation is exactly solved with the
modified Numerov method and the ingoing wave
boundary condition imposed inside (about 2 fm) the capture barrier. 
The fusion cross section $\sigma_{fus}$ is calculated taken into account the 
identity of the interacting nuclei and the parity of the wave function 
for the relative motion (only even partial waves $L$ are included here),
i.e., $\sigma_{fus} = \pi \hbar^2/(2 \mu E) \sum_{L}
(2L+1)(1+\delta_{1,2})P_L$, where $\mu$ is the asymptotic reduced
mass, $E$ is the incident energy in the total center-of-mass
reference frame and $P_L$ is the partial tunneling probability.

Figure \ref{SFACT} shows the S-factor as a function of the incident
energy in the center-of-mass reference frame. For a better
presentation, the experimental data of each set\cite{Exp} are binned 
into $\Delta E = 0.5$ MeV energy intervals. 
In this figure the following features can be
observed:
\begin{romanlist}[(ii)]
\item the molecular adiabatic potential of Fig.
\ref{POTINERT}a correctly (thick and thin solid curves) explains
the measured data, in contrast to either the results
obtained with the BW potential (dotted curve) or the very recent 
calculations within the Fermionic Molecular 
Dynamics (FMD) approach\cite{Neff} (dashed curve). Since the width of the 
barrier decreases for the molecular adiabatic potential of Fig. \ref{POTINERT}a, 
it produces larger fusion cross sections than those arising from the 
shallower KNS and BW potentials. 
\item the use
of the cranking mass parameter of Fig. \ref{POTINERT}b notably
affects the low energy S-factor, which is revealed by the
comparison between the thick and thin solid curves. It starts
reducing the S-factor around 7-8 MeV energy region and produces a
local maximum around 4.5 MeV. At the lowest incident energies (below
4 MeV) the S-factor is suppressed by a factor of five compared to
that arising from a constant reduced mass. The peak in the S-factor 
is due to an increase of the fusion cross section, which is 
caused by the resonant behavior of the collective radial wave 
function\cite{Alexis1}. 
\end{romanlist}

\section{Concluding remarks and outlook}

The adiabatic molecular picture very well explains the available experimental data 
for the S-factor of $^{16}$O + $^{16}$O. The collective radial cranking mass 
causes a relevant peak in the S-factor excitation function, although the pairing 
correlation (neglected in the calculation) may somewhat reduce the magnitude of this bump. It is highly desirable to have fusion cross sections, measured around 4-5 MeV, to verify the existence of this resonant structure in the S-factor. The collective mass surface is very important for the reaction dynamics, and its effect on fusion of heavy-ions should be investigated systematically. This can be significant for a better understanding of reactions forming superheavy elements. Works are in progress for understanding the molecular resonance structures in the S-factor excitation function of the challenging system $^{12}$C + $^{12}$C.

\section*{Acknowledgements}

This work was supported by the Joint Institute for Nuclear Astrophysics (JINA) through 
grant NSF PHY 0216783, and by an ARC Discovery grant.


\end{document}